\documentclass[twocolumn]{aastex63}
\usepackage{graphicx}
\usepackage{amsmath}
\usepackage{booktabs}
\usepackage{multirow}
\usepackage{float}
\usepackage{lineno}
\usepackage{subfigure}
\usepackage{subfloat}
\usepackage{overpic}
\usepackage{ulem}
\usepackage{longtable}
\usepackage[flushleft]{threeparttable}
\usepackage{hyperref}
\usepackage[hyphenbreaks]{breakurl}
\usepackage{CJKutf8}
\usepackage[normalem]{ulem}

\shorttitle{General Trigger Classification Framework}
\shortauthors{Zhang et al.}

\begin{document}
\begin{CJK*}{UTF8}{gbsn}

\title{\texttt{LUNCH}: A Lightweight Unified Deep-Learning Framework for General Transients Classification in High-Energy Time-Domain Astronomy}

\correspondingauthor{Ren-Zhou Gui, Shao-Lin Xiong, Xiao-Bo Li}
\email{rzgui@tongji.edu.cn, xiongsl@ihep.ac.cn, lixb@ihep.ac.cn}

\def\Tongji{College of Electronic and Information Engineering, Tongji University, Shanghai 201804, China}

\def\HighEnergy{State Key Laboratory of Particle Astrophysics, Institute of High Energy Physics, Chinese Academy of Sciences, Beijing 100049, China}

\def\Guokeda{University of Chinese Academy of Sciences, Beijing 100049, Beijing, China}

\def\jinggangshan{Department of Physics, Jinggangshan University, Jiangxi Province, Ji'an 343009, China}

\def\Nanjing{School of Astronomy and Space Science, Nanjing University, Nanjing 210023, China}

\def\Guangxi{Guangxi Key Laboratory for Relativistic Astrophysics, Nanning 530004, China}

\def\XinanJiaoTong{School of Computing and Artificial Intelligence, Southwest
Jiaotong University, Chengdu 611756, China}

\def\XiAnJiaoTong{Department of Nuclear Science and Technology, School of Energy and Power Engineering, Xi’an Jiaotong University, Xi’an 710049, China}

\def\HeBeiNormal{College of Physics and Hebei Key Laboratory of Photophysics Research and Application, Hebei Normal University, Shijiazhuang, Hebei 050024, China}

\def\GuiZhouNormal{School of Physics and Electronic Science, Guizhou Normal University, Guiyang 550001, China}

\def\Dezhou{School of Computer and Information, Dezhou University, Dezhou 253023, China}

\def\BeijingNormal{Department of Astronomy, Beijing Normal University, Beijing 100875, China}

\def\HuazhongNormal{Institute of Astrophysics, Central China Normal University, Wuhan 430079, HuBei, China}

\def\FuDan{College of Computer Science and Artificial Intelligence, Fudan University, 200433}

\author[0000-0002-8097-3616]{Peng Zhang}
\affiliation{\Tongji}
\affiliation{\HighEnergy{}}

\author[0009-0008-8053-2985]{Chen-Wei Wang}
\affiliation{\HighEnergy{}}

\author[0009-0002-6411-8422]{Zheng-Hang Yu}
\affiliation{\HighEnergy{}}

\author{Ren-Zhou Gui}
\affiliation{\Tongji}

\author[0000-0002-4771-7653]{Shao-Lin Xiong}
\affiliation{\HighEnergy{}}

\author{Xiao-Bo Li}
\affiliation{\HighEnergy{}}

\author[0000-0003-0274-3396]{Li-Ming Song}
\affiliation{\HighEnergy{}}

\author{Shi-Jie Zheng}
\affiliation{\HighEnergy{}}

\author{Xiao-Yun Zhao}
\affiliation{\HighEnergy{}}

\author{Yue Huang}
\affiliation{\HighEnergy{}}

\author[0000-0001-8664-5085]{Wang-Chen Xue}
\affiliation{\HighEnergy{}}

\author[0000-0003-3789-6368]{Ya-Qi Wang}
\affiliation{\Tongji}

\author[0000-0001-5122-8468]{Long-Bo Han}
\affiliation{\FuDan}

\author[0009-0004-1887-4686]{Jia-Cong Liu}
\affiliation{\HighEnergy{}}

\author[0009-0001-7226-2355]{Chao Zheng}
\affiliation{\HighEnergy{}}

\author[0009-0006-5506-5970]{Wen-Jun Tan}
\affiliation{\HighEnergy{}}

\author[0000-0001-9217-7070]{Sheng-Lun Xie}
\affiliation{\HighEnergy{}}
\affiliation{\HuazhongNormal{}}

\author[0000-0002-6540-2372]{Ce Cai}
\affiliation{\HeBeiNormal{}}

\author[0000-0001-5348-7033]{Yan-Qiu Zhang}
\affiliation{\GuiZhouNormal{}}

\author[0009-0004-5711-2692]{Hao-Xuan Guo}
\affiliation{\XiAnJiaoTong}

\author[0009-0008-5068-3504]{Yue Wang}
\affiliation{\HighEnergy{}}

\author{Yang-Zhao Ren}
\affiliation{\HighEnergy{}}
\affiliation{\XinanJiaoTong{}}

\begin{abstract}
The increasing data volume of high‑energy space monitors necessitates real‑time, automated transient classification for multi‑messenger follow‑up. 
Conventional methods rely on empirical features like hardness ratios and reliable localization, which are not always precisely available during early detection. 
We developed the Lightweight Unified Neural Classifier for High‑energy Transients (\texttt{LUNCH})—an end‑to‑end deep‑learning framework that performs general transient classification directly from raw multi‑band light curves, eliminating the need for background subtraction or source localization.
Its dual‑scale architecture fuses long‑ and short‑scale temporal evolution adaptively. 
Evaluated on 15 years of Fermi/GBM triggers, the optimal model achieves 97.23\% accuracy when trained on complete energy spectra. 
A lightweight version using only three broad energy bands retains 95.07\% accuracy, demonstrating that coarse spectral information fused with temporal context enables robust discrimination. 
The system significantly outperforms the GBM in-flight classifier on three months of independent test data. 
Feature visualization reveals well‑separated class clusters, confirming physical interpretability. \texttt{LUNCH} combines high accuracy, low computational cost, and instrument‑agnostic inputs, offering a practical solution for real‑time in-flight processing that enables timely triggers for immediate multi‑wavelength and multi‑messenger follow‑up observations in future time‑domain missions.
\end{abstract}

\keywords{Gamma-ray astronomy (628), Gamma-ray bursts (629), 
High energy astrophysics (739), Convolutional neural networks (1938), Astronomy data analysis (1858)}

\section{Introduction}
\label{sect:intro}
Time-domain astronomy has revolutionized our understanding of the dynamic universe by systematically capturing transient and variable phenomena across the electromagnetic spectrum.  
Modern wide-field monitoring instruments such as the Gravitational Wave High-energy Electromagnetic Counterpart All-sky Monitor 
(GECAM-A/B \citet{GECAM}, GECAM-C \citet{HEBS_INS_Zhang2023}, and GECAM-D \citet{GTM_INS_wang2024,GTM_INS_Feng2024}), the Hard X-ray Modulation Telescope (\textit{Insight}-HXMT/HE, \citet{HXMT}), the \textit{Fermi} Gamma-ray Burst Monitor (GBM, \citet{Fermi_GBM}), and the Space-based multi-band astronomical Variable Objects Monitor (SVOM/GRM, \citet{SVOM_GRM}) conduct continuous, all-sky surveys in the high-energy regime. 
The increasingly sensitive trigger search algorithms generate a stream of candidate transients~\citep{trig_search_cai_2021,trig_search_cai_2025,GECAM_ground_trig_search}. 
The events thus identified represent a diverse astrophysical, encompassing cosmological gamma-ray bursts (GRBs), magnetar-driven soft gamma repeater (SGR) bursts, atmospheric terrestrial gamma-ray flashes (TGFs), solar flares, and various particle-induced phenomena. 
The rapid and reliable classification of these triggers is paramount, as it serves as the critical gateway to enabling timely multi-wavelength and multi-messenger follow-up, efficiently prioritizing limited observational resources, and ultimately unlocking the full discovery potential of both current and next-generation high-energy missions.
\par
Traditional trigger classification pipelines rely on manually engineered empirical features—such as source localization, hardness ratios, and geomagnetic information—typically implemented through Bayesian or threshold-based decision rules that depend on external priors and instrument response models \citep{trig_classify_1, Fermi_GBM, trig_classify_3, trig_classify_gecam}. 
A core limitation of these approaches is they strongly depend on accurate background estimation (for deriving reliable hardness ratios) and on precise localization (for excluding known sources). 
However, these requirements are often unmet in practice, especially for instruments with limited or no intrinsic localization capability (e.g., HXMT/HE-CsI, GRID \citep{GRID_INS_Wen2019}, GECAM-D, SVOM/GRM, Fermi/GBM) or those operating in complex background environments. 
Moreover, the frequent reliance on auxiliary external data—such as geomagnetic or solar activity information to discriminate events like solar flares or geomagnetic storms—introduces two critical constraints: first, classification reliability degrades when such data are unavailable;
second, the need to integrate these external sources inherently rules out real-time trigger classification. 
\par
Driven by these limitations, feature-engineered methods have motivated a shift towards data-driven approaches. 
Deep learning (DL) has emerged as a powerful tool in astronomy due to its capacity to automatically learn hierarchical, abstract representations from raw data~\citep{AI_astro_Djorgovski2022,AI_astro_Lieu2025,AI_astro_Wang2025}. 
Its application to time-domain data, particularly for classifying gamma-ray transients, has shown considerable promise. 
Supervised convolutional neural networks (CNNs) have been used with high accuracy to distinguish GRBs from background, utilizing multi-band light curves or count maps~\citep{DL_search_grb_zhangpeng_1, DL_search_grb_zhangpeng_2}. 
\citet{DL_GRB_classify_chen_2025apjs} employed a CNN–based method for GRB classification. 
This method resolves the duration overlap issue by utilizing temporal-spectral count maps, distinguishes between short and long GRBs, and confirms their associations with divergent types of supernovae and kilonovae, achieving higher accuracy and robustness than conventional approaches. 
Unsupervised methods, such as auto-encoders for anomaly detection, have successfully identified GRBs missed by standard catalogs \citep{DL_search_grb_Parmiggiani_2023}. 
\par
In the specific context of trigger classification, recent studies have applied DL models to data from individual missions.
Using the GECAM-B data, \cite{trig_classify_DL_0} migrated transformer model to achieve 89\% accuracy classification (GRBs, SFs, SGRs and PARTICLEs) from the multi-energy light curves of multi-detector and the incident angle information of known sources. 
\cite{trig_classify_DL_1} presented two deep learning-based classifiers, employing convolutional and recurrent neural networks, which categorize Fermi-GBM gamma-ray transients (GRBs, TGFs, solar flares, and SGRs) from background-subtracted light curves, achieving an overall accuracy of 93\%.
However, a critical gap remains: these DL models are typically trained and optimized for a specific instrument's data characteristics. 
They lack true generality—that is, the ability to perform robust, high-fidelity classification across different satellite missions without retraining or significant adaptation. 
This capability is increasingly vital in the era of multi-messenger, multi-instrument time-domain astronomy.
\par
Despite these advances, a critical limitation persists: existing DL classifiers are typically tailored to a single instrument’s data characteristics and pre-processing pipelines (e.g., background subtraction) and often depend on instrument-specific inputs like localization estimates. 
This instrument specificity limits cross-mission generalization, undermining robustness in multi-instrument, multi-messenger observational contexts where trigger data characteristics vary substantially. 
For time-domain astronomy to fully leverage the growing volume of diverse trigger streams, classification frameworks must be both instrument-agnostic and robust to systematic differences among datasets.
\par
In this work, we address this gap by introducing the Lightweight Unified Neural Classifier for High‑energy Transients (\texttt{LUNCH}). 
It is a general deep learning framework for astronomical trigger classification that operates directly on universally available multi-energy-band light curves. 
Our model employs a dual-scale architecture to capture both macroscopic temporal evolution and fine-scale variability, and incorporates an attention-based fusion module to dynamically integrate multi-scale features.
Unlike many existing approaches, this framework does not depend on background-subtracted count rates, localization information, or instrument-specific priors, enabling robust performance across diverse observational regimes. 
We validate our approach on the extensive \textit{Fermi}/GBM trigger catalog, achieving state-of-the-art classification accuracy across five major transient classes—GRBs, SGRs, TGFs, solar flares, and particle events—and demonstrate its practical suitability for automated data pipelines and real-time transient response systems. 
\par
The remainder of this paper is organized as follows. Section~\ref{sec:dataset} describes the dataset and pre-processing. Section~\ref{sec:method} details the proposed model architecture and training strategy. Section~\ref{sec:results} presents comprehensive evaluations, including ablation studies and feature analysis. Section~\ref{sec:discussion} discusses implications and future directions.

\begin{figure*}
\centering
\begin{minipage}[b]{0.46\textwidth}
    \centering
    \includegraphics[width=\linewidth]{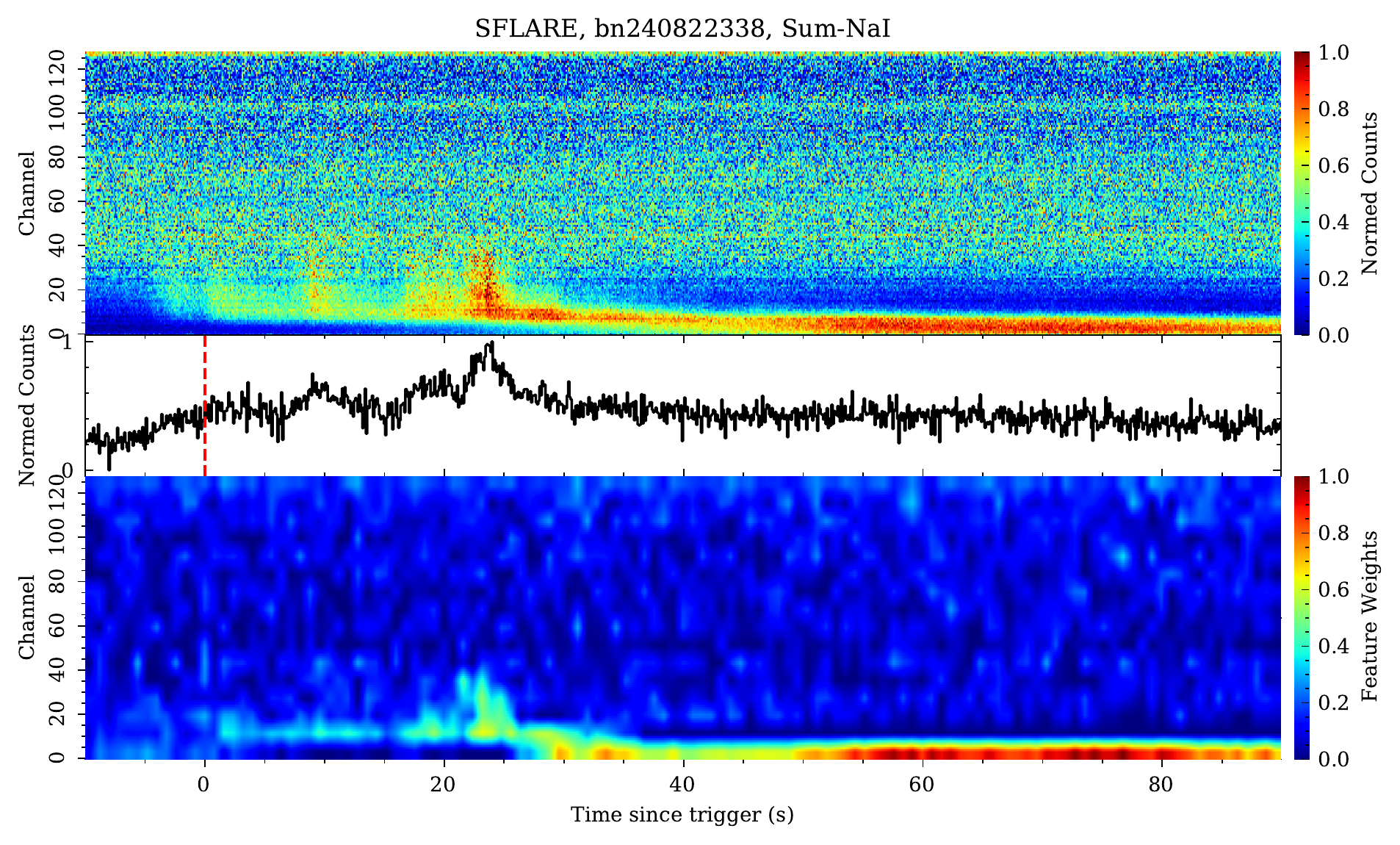}
\end{minipage}
\begin{minipage}[b]{0.46\textwidth}
    \centering
    \includegraphics[width=\linewidth]{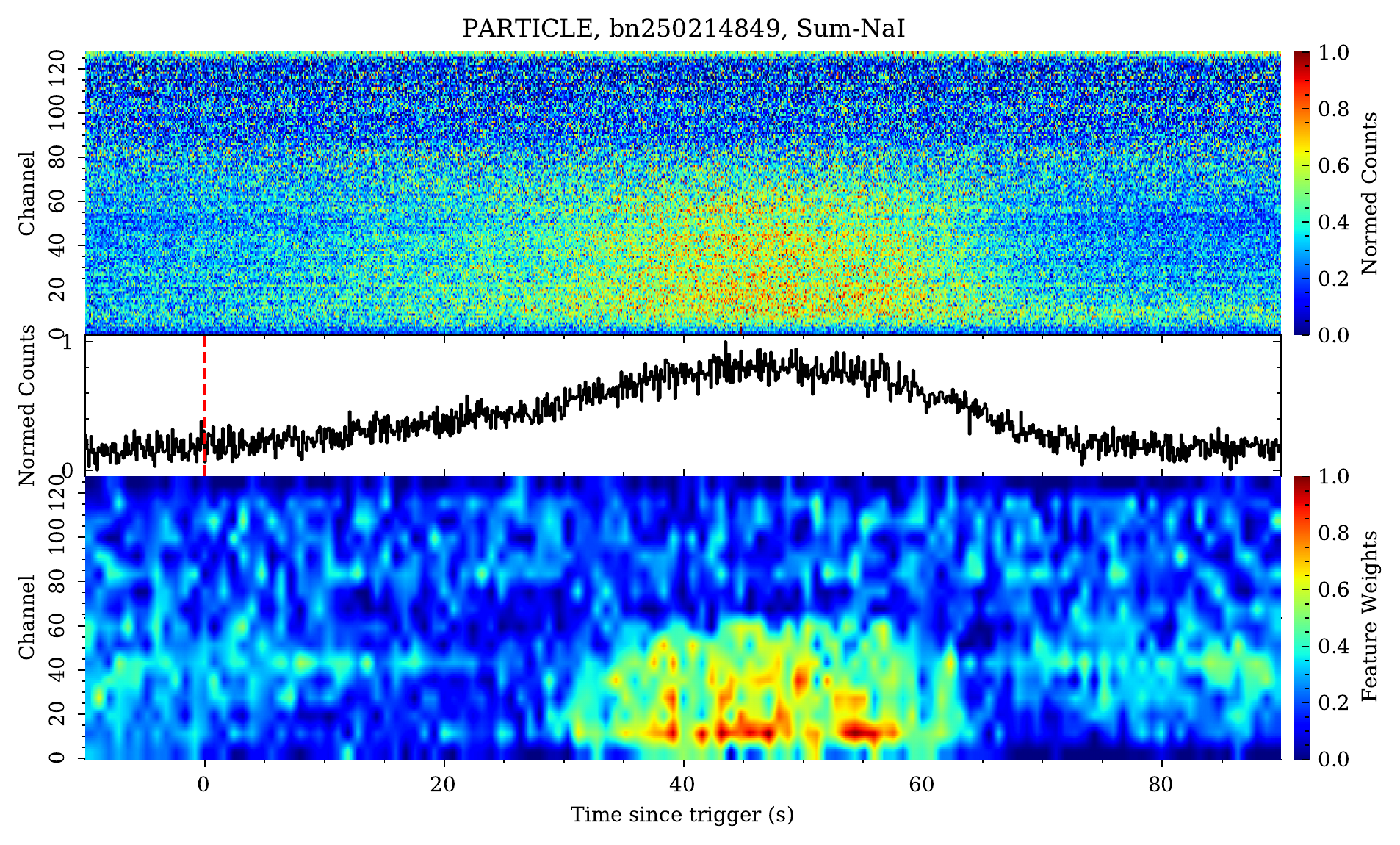}
\end{minipage}
\hspace{0.001\textwidth}
\begin{minipage}[b]{0.46\textwidth}
    \centering
    \includegraphics[width=\linewidth]{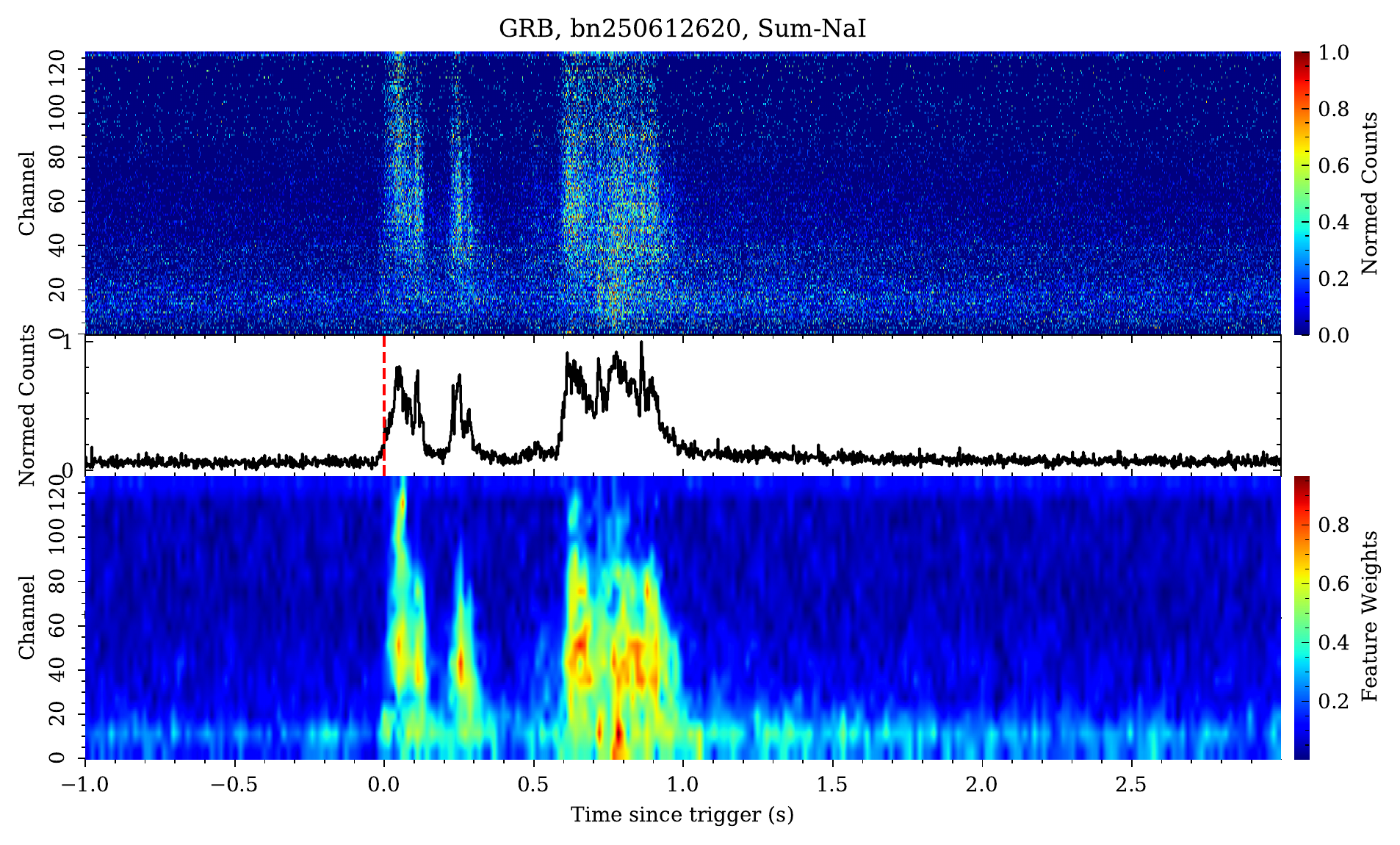}
\end{minipage}

\centering
\begin{minipage}[b]{0.46\textwidth}
    \centering
    \includegraphics[width=\linewidth]{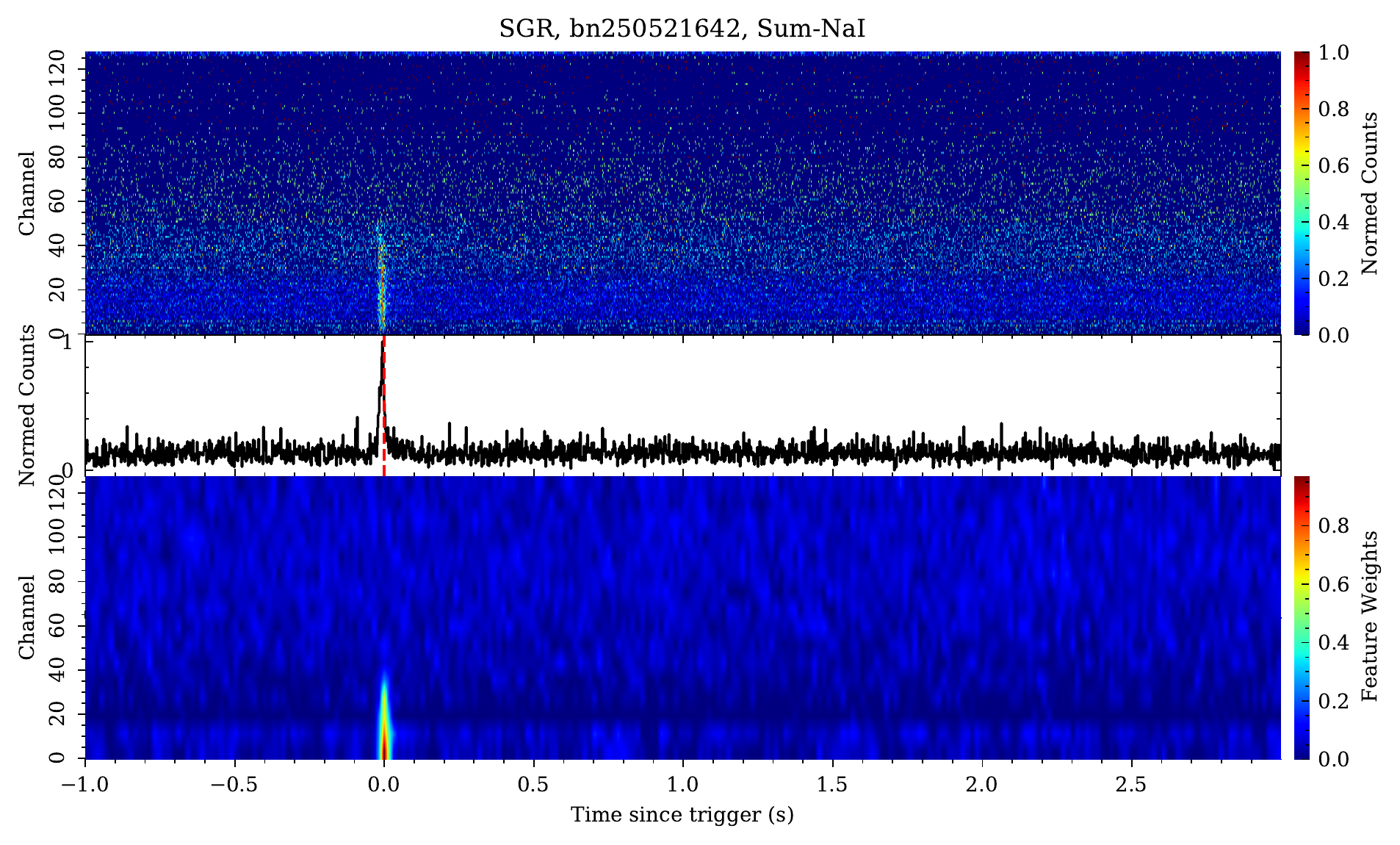}
\end{minipage}
\hspace{0.001\textwidth}
\begin{minipage}[b]{0.46\textwidth}
    \centering
    \includegraphics[width=\linewidth]{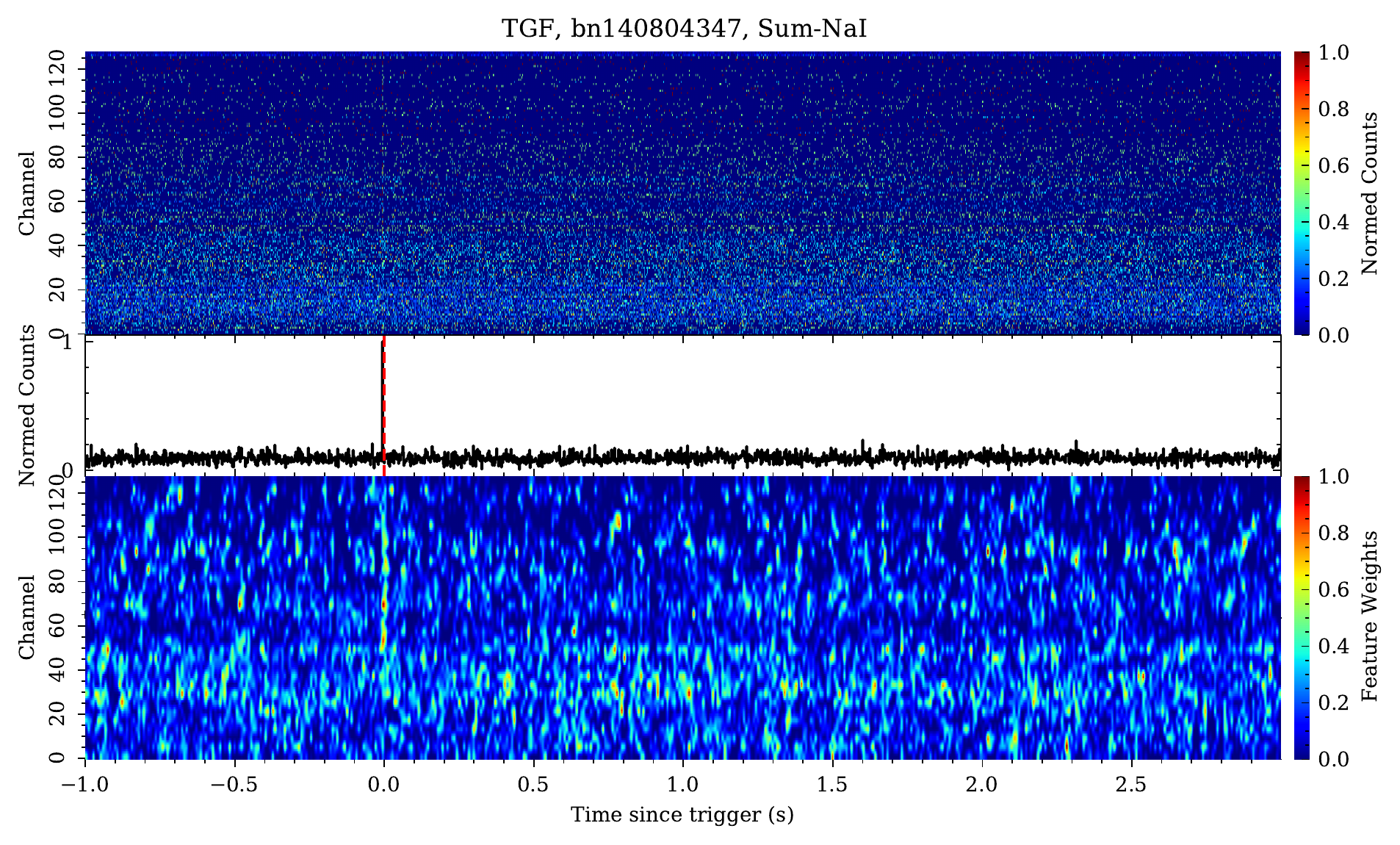}
\end{minipage}
\caption{
Representative samples of the five trigger classes with Grad-CAM feature visualizations.
For each event, the top panel shows the normalized multi-channel (128 energy bins) count map, the middle panel shows the summed light curve across all energy channels, and the bottom panel displays the Grad-CAM heatmap, highlighting the spatio-temporal features (time-energy bins) that most influenced the model's classification. 
}
\label{fig:trig_samples}
\end{figure*}

\begin{deluxetable*}{cccccc}[h]
\label{table:dataset}
\tablecaption{Class Distribution of the \textit{Fermi}/GBM Trigger Dataset}
\tablehead{Partition & GRB & SFLARE & SGR & TGF & PARTICLE}
\startdata
Training set & 1670 & 1279 & 313 & 906 & 701\\
Validation set & 556 & 426 & 105 & 302 & 234\\
Test set & 557 & 426 & 104 & 302 & 234\\
\enddata
\end{deluxetable*}

\begin{figure*}[htbp]
\centering
\includegraphics[width=\textwidth]{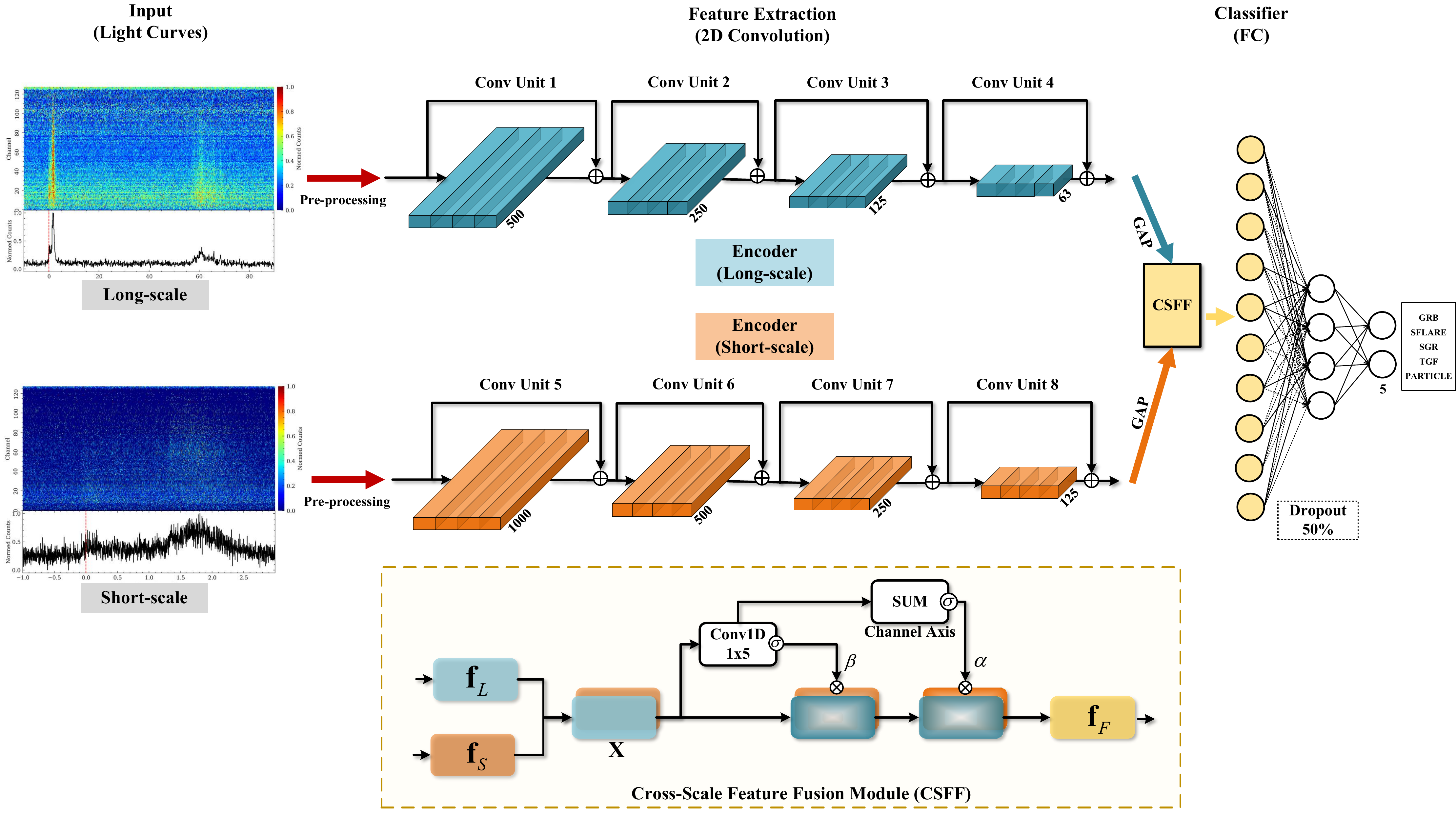}
\caption{
\textbf{Schematic of the proposed dual-scale neural network architecture for general trigger classification.}
The model processes input light curves through two parallel pathways: a long-scale encoder (blue) and a short-scale encoder (orange), each composed of four consecutive 2D convolutional units (Conv Units 1-4 and 5-8, respectively) for hierarchical feature extraction. 
The extracted feature vectors, $\mathbf{f}_L$ and $\mathbf{f}_S$, are fed into the central Cross-Scale Feature Fusion Module (CSFF). The zoomed-in view details the CSFF's operation: a 1D convolutional layer models cross-scale interactions, generating attention maps which adaptively weight and combine $\mathbf{f}_L$ and $\mathbf{f}_S$ to produce a unified representation $\mathbf{f}_F$. This fused feature is aggregated via global average pooling (GAP) and finally classified by a fully-connected (FC) layer with dropout. This end-to-end architecture enables the joint analysis of complementary temporal information. 
}
\label{fig:model}
\end{figure*}

\section{Dataset}\label{sec:dataset}
\subsection{Instrumentation: Fermi Gamma-ray Burst Monitor}
The Fermi Gamma-ray Burst Monitor (GBM) is a dedicated instrument onboard the Fermi Gamma-ray Space Telescope, designed for continuous, all-sky monitoring in the energy range of approximately 8\,keV to 40\,MeV \citep{Fermi_GBM}. The instrument comprises 12 semi-directional sodium iodide (NaI) scintillation detectors, sensitive to photons between 8 and 1000\,keV, and two bismuth germanate (BGO) detectors, which cover the higher energy band from 200\,keV to 40\,MeV. The NaI detectors are oriented to provide nearly complete, unocculted sky coverage, while the two BGO detectors are mounted on opposite sides of the spacecraft to ensure broad angular sensitivity. 
Since its activation on 12 July 2008, the GBM has provided a continuous data stream, 
establishing a facility for high-energy, time-domain astrophysics. 
Its comprehensive trigger catalog and extensive observational timeline make it an ideal testbed for validating the proposed general trigger classification framework. 

\subsection{Event Selection}
We utilize the complete, publicly available GBM trigger catalog\footnote{\url{https://heasarc.gsfc.nasa.gov/W3Browse/fermi/fermigtrig.html}} spanning from 12 July 2008 through 31 August 2025, encompassing the instrument's full operational period with calibrated instrumental response. 
The catalog documents 11,911 manually checked trigger events, which are distributed according to the established astrophysical and instrumental taxonomy as follows: Gamma-Ray Bursts (GRB; 4,075 events), Solar Flares (SFLARE; 2,534), Soft Gamma Repeaters (SGR; 671), Terrestrial Gamma-Ray Flashes (TGF; 1,558), and particle-induced events.
Particle-induced events are further subdivided into Distance Particle events (DISTPAR; 124) and Local Particle events (LOCLPAR; 1,771) based on their inferred origin.
To alleviate the significant class imbalance while maintaining physical interpretability, we adopt a consolidated five-category classification scheme: GRB, SFLARE, SGR, TGF, and a composite PARTICLE class. The PARTICLE category amalgamates the DISTPAR and LOCLPAR subclasses. 
This consolidation is motivated by their common origin as charged particle interactions with the spacecraft or instrument, and by the small sample size of DISTPAR events relative to the other major categories.
This grouping ensures adequate sample sizes for robust machine learning model training while preserving the essential physical distinction between astrophysical transients and instrumental/background phenomena.

\subsection{Data Extraction and Pre-processing}
\label{subsec:data_extraction}
For each trigger event, we employ a multi-detector, multi-scale data extraction strategy to capture the broad spectral coverage and characteristic temporal variability of astrophysical transients~\citep{trig_classify_DL_1,trig_search_cai_2025}. 

Temporal Scales: We extract data at two complementary temporal resolutions:
\begin{itemize}
    \item \textbf{Long-scale:} 0.1\,s resolution, covering a window from $T_0 - 10$\,s to $T_0 + 90$\,s around the trigger time $T_0$, designed to capture the overall light curve morphology and any extended emission components.
    \item \textbf{Short-scale:} 2\,ms resolution, covering a window from $T_0 - 1$\,s to $T_0 + 3$\,s, optimized to resolve rapid variability and fine-structure features critical for distinguishing fast transients like TGFs and short GRBs.
\end{itemize}

Data are extracted from the Time-Tagged Event (TTE) files\footnote{\url{https://heasarc.gsfc.nasa.gov/FTP/fermi/data/gbm/triggers/}} associated with each trigger, which provide 128-channel spectral resolution for both NaI and BGO detectors, thereby enabling the reconstruction of spectral evolution without applying model-dependent corrections.
To maximize the signal-to-noise ratio and improve spectral characterization, we generate composite light curves by combining data from all 12 NaI detectors or both BGO detectors for each event. 
This approach thus preserves the intrinsic spectral-temporal correlations that are essential for classification.
\par
A critical pre-processing step is the per-energy-channel standardization of the light curves. 
We scale the photon counts in each spectral channel independently to the range [0, 1] based on the minimum and maximum values observed across the training set for that channel. 
Empirically, this method proves superior to conventional $z$-score normalization because it preserves the critical relative inter-channel flux ratios. 
This preservation maintains the integrity of the incident energy spectrum, which in turn serves as a highly discriminative feature for classification algorithms. 
\par
We implemented a stringent selection criterion to ensure the highest purity of training labels and prevent contamination from ambiguous or misidentified events. 
Events from the GBM trigger catalog were retained only if their catalog \texttt{Reliability'} flag indicated 100\% classification confidence, following the approach of \citet{trig_classify_DL_1}.
This approach guarantees the highest possible fidelity for the human-curated ground-truth labels used in supervised learning. 
After applying this reliability-based selection and quality filters to exclude events with incomplete or corrupted data, the final curated dataset comprises 8,115 events. 
The dataset was partitioned into training, validation, and test subsets (6:2:2 ratio) using stratified random sampling to preserve the original class distribution, thereby alleviating the effects of class imbalance during model development and evaluation.
Table~\ref{table:dataset} presents the detailed composition of the final dataset, including class-wise sample counts for each subset.

\section{Method}\label{sec:method}

\subsection{Architecture of Neural Networks}
\label{subsec:architecture}
The overall architecture of the proposed neural network is designed as a dual-pathway encoder, as illustrated in Figure~\ref{fig:model}. 
Its primary function is to process gamma-ray transient light curves at two distinct temporal resolutions concurrently, thereby learning complementary representations from both their macroscopic evolution and fine-scale variability.
The input multi-channel light curves first undergo a pre-processing stage, after which the processed data flow through two parallel branches: a long-scale encoder (color-coded in blue in the schematic) and a short-scale encoder (color-coded in orange). 
Each encoder is implemented as a sequence of four dedicated 2D convolutional blocks (denoted as Conv Unit 1-4 for the long-scale and Conv Unit 5-8 for the short-scale path). 
Each Conv Unit comprises four convolutional layers, each using 3×3 kernels. 
The first layer employs a stride of 2 for down‑sampling, while the subsequent layers use a stride of 1.
These blocks perform hierarchical feature extraction, transforming the raw input into high-level, abstract feature vectors \(\mathbf{f}_L\) and \(\mathbf{f}_S\) for the long and short-scale paths, respectively.
\par
These feature vectors are then integrated by the central Cross-Scale Feature Fusion Module (CSFF). 
This module dynamically combines \(\mathbf{f}_L\) and \(\mathbf{f}_S\) using a cross-attention strategy (detailed in the following subsection) to produce a unified, context-aware feature representation \(\mathbf{f}_F\). 
The fused feature vector \(\mathbf{f}_F\) is then aggregated via a Global Average Pooling (GAP) layer and fed to the final classification head. 
The classifier consists of a fully-connected (FC) layer, preceded by a Dropout layer (with a rate of 50\%) for regularization, and ultimately produces a five-dimensional output corresponding to the class probabilities.
This end-to-end design enables the model to learn discriminative patterns directly from the raw temporal data across multiple scales. 
We optimized the network architecture by performing a hyperparameter search: the number of convolutional filters was selected from {16, 32, 64}, and the number of neurons in the FC layer from {8, 16, 32, 64}.

\subsection{Cross-Scale Feature Fusion Module (CSFF)}
\label{subsec:fusion_module}

To effectively integrate the complementary information extracted from the long-scale and short-scale pathways, we designed a dedicated Cross-Scale Feature Fusion Module (CSFF). This module takes as input the feature vectors output by the two pathway encoders, denoted as $\mathbf{f}_L \in \mathbb{R}^{B \times D}$ and $\mathbf{f}_S \in \mathbb{R}^{B \times D}$, where $B$ is the batch size and $D$ is the feature dimension (default $D=64$). The core objective is to learn a set of data-dependent attention weights that govern how features from each temporal scale should be adaptively combined for optimal classification.

The fusion process, illustrated in the bottom of Figure~\ref{fig:model}, proceeds as follows. First, the two feature vectors are concatenated along a new dimension to form a structured feature tensor:

\begin{equation}
\mathbf{X} = [\mathbf{f}_L, \mathbf{f}_S] \in \mathbb{R}^{B \times 2 \times D},
\label{eq:concat}
\end{equation}

\noindent where the second dimension explicitly represents the two scales (long and short). A 1D convolutional layer with a kernel size of $k$ (default $k=5$) is then applied to this tensor to model local interactions between corresponding feature elements across scales, generating an intermediate attention map:

\begin{equation}
\mathbf{A} = \text{Conv1D}_{k}(\mathbf{X}) \in \mathbb{R}^{B \times 2 \times D}.
\label{eq:conv_att}
\end{equation}

The module computes attention at two hierarchical levels. A \textit{scale-level attention} vector $\boldsymbol{\alpha} \in \mathbb{R}^{B \times 2}$ is derived by global pooling of $\mathbf{A}$ along the feature dimension followed by a sigmoid activation:

\begin{equation}
\boldsymbol{\alpha} = \sigma\left(\sum_{d=1}^{D} \mathbf{A}[:,:,d]\right),
\label{eq:scale_att}
\end{equation}

\noindent where $\sigma(\cdot)$ denotes the sigmoid function. This vector captures the global importance of the entire feature set from each scale for the current input sample. Concurrently, a finer-grained \textit{element-wise attention} map is obtained via another sigmoid activation:

\begin{equation}
\mathbf{\beta} = \sigma(\mathbf{A}) \in \mathbb{R}^{B \times 2 \times D}.
\label{eq:elem_att}
\end{equation}

\noindent This map modulates the contribution of each individual feature dimension from each scale.

The final fused feature vector $\mathbf{f}_F \in \mathbb{R}^{B \times D}$ is computed as a gated combination of the original features, weighted by both attention mechanisms:

\begin{equation}
\mathbf{f}_F = \sum_{i=1}^{2} \left( \mathbf{X}[:,i,:] \odot \mathbf{\beta}[:,i,:] \odot \boldsymbol{\alpha}[:,i] \right),
\label{eq:final_fusion}
\end{equation}

\noindent where $\odot$ denotes element-wise multiplication, and $\boldsymbol{\alpha}_s[:,i]$ is broadcast along the feature dimension. The summation over the scale dimension ($i=1$ for long-scale, $i=2$ for short-scale) yields the final, context-aware fused representation.

This dual-attention design enables the module to dynamically emphasize the more reliable temporal scale globally while simultaneously performing fine-grained, feature-wise recalibration. The resulting fused representation $\mathbf{f}_F$ thus incorporates complementary information from both pathways in a data-adaptive manner, providing a robust input to the final classification layer.

\subsection{Model Training and Analysis}
\label{subsec:training}
We employed the Focal Loss \citep{focal_loss} as the loss function to alleviate the intrinsic class imbalance in the astrophysical trigger dataset and enhance the model's focus on challenging samples. The loss is defined as:

\begin{equation}
    \mathcal{L}_{\text{focal}} = \frac{1}{N} \sum_{i=1}^{N} (1 - p_{t,i})^{\gamma} \cdot \mathcal{L}_{\text{CE}}(y_i, \hat{y}_i),
\end{equation}

\noindent where $N$ is the batch size, $p_{t,i}$ is the model's predicted probability for the true class of sample $i$, $\gamma$ is a focusing parameter (set to 2.0), and $\mathcal{L}_{\text{CE}}$ is the standard cross-entropy loss. The Focal Loss dynamically down-weights the contribution of well-classified easy examples, allowing the training to concentrate on hard, misclassified events. 
The model was optimized using the Adam optimizer \citep{adam_optimizer} with an initial learning rate of $5 \times 10^{-4}$ and a weight decay coefficient of $1 \times 10^{-3}$ for $\ell_2$ regularization. A scheduler was utilized to dynamically adjust the learning rate, reducing it by a factor of 0.5 whenever the validation loss failed to improve for 5 consecutive epochs, with a minimum learning rate of $1 \times 10^{-8}$.
\par
The network was trained with a batch size of 32. To prevent overfitting and ensure the selection of the most generalizable model, an early stopping mechanism with a patience of 20 epochs was implemented. Training was halted if no improvement in the validation loss was observed for 20 consecutive epochs. The model state corresponding to the epoch with the lowest validation loss was retained as the final model. 
To ensure reproducibility, all experiments were conducted using PyTorch 1.12.1 with fixed random seeds. Hardware acceleration was provided by an NVIDIA 4090 GPU, with typical training duration ranging from 15 to 45 minutes per model configuration depending on architectural complexity.

\subsection{Feature Visualization}
\label{subsec:feature_viz}
To interpret the model's internal representations and classification decisions, we employed two complementary visualization techniques: Uniform Manifold Approximation and Projection (UMAP) for analyzing the global structure of the learned feature space, and Gradient-weighted Class Activation Mapping (Grad-CAM) for localizing discriminative regions within the input data.
These two methods are well-established techniques for feature visualization in astronomy~\citep{Classify_GRB_Dimple_2023apj,DL_search_grb_zhangpeng_1,Waterfalls_classify_GRB_2025apj,DL_search_grb_zhangpeng_2}.

\textit{UMAP for Visualizing High-dimensional Features.} 
We utilized the UMAP algorithm \citep{UMAP} to project the high-dimensional feature vectors—extracted from the layer preceding the final classifier in our optimal model—onto a two-dimensional plane. 
UMAP constructs a low-dimensional representation that preserves both the local and global topological structure of the high-dimensional data manifold. 
This non-linear projection enables an intuitive visual assessment of whether the model learns semantically separable representations for different trigger classes.
A clear separation of clusters in the 2D UMAP plot would indicate that the model’s latent space effectively encodes distinctive features for each class, providing a geometric explanation for its classification performance. 
UMAP hyper-parameters, including the number of neighbors and the minimum distance, were systematically explored across a broad range (2 $\leq$ \texttt{n\_neighbors} $\leq$ 50; 0.01 $\leq$ \texttt{min\_dist} $\leq$ 0.99) to ensure stability and robustness of the visualization. 
The final configuration (\texttt{n\_neighbors} = 3 and \texttt{min\_dist} = 0.05) was selected to achieve a balanced preservation of global structure and local neighborhood fidelity, which is particularly important for interpreting high-dimensional transient light-curve representations.

\textit{Grad-CAM for Interpreting Classification Decisions.} 
To identify the specific temporal and spectral regions in the input light curves that were most critical for the model's predictions, we applied the Grad-CAM technique \citep{model_grad_cam}. 
For a given input and its predicted class, Grad-CAM computes the gradients of the class score with respect to the activations in the final convolutional layer. 
These gradients are globally average-pooled to obtain neuron importance weights, which are then used to generate a coarse localization heatmap over the input temporal-spectral grid.
This heatmap highlights the areas (i.e., specific time bins and energy channels) that the model attended to when making a particular classification. 
Analyzing these Grad-CAM visualizations helps validate that the model’s decisions are based on physically meaningful features (e.g., the peak of a GRB or the duration of a solar flare) rather than spurious correlations, thereby enhancing the trustworthiness and interpretability of the deep learning model.

\section{Results}
\label{sec:results}

\subsection{Typical Samples and Model Interpretation}
\label{subsec:results_cam}
Figure~\ref{fig:trig_samples} shows representative events from each class, alongside the corresponding Grad-CAM visualizations generated by our optimal model. 
The Grad-CAM visualizations reveal how the proposed framework identifies discriminative temporal-spectral patterns characteristic of each trigger class. 
We observe distinct and physically interpretable activation patterns: GRBs and TGFs exhibit compact, high-intensity features localized in time, while SFLARE and PARTICLE events show broader or irregular activations, consistent with their extended temporal profiles or non-astrophysical origins.
These results indicate that the model bases its decisions on meaningful spatio-temporal structures (time-energy bins) rather than spurious correlations.

\subsection{Single-Scale Performance}
\label{subsec:results_single}
We next evaluate the classification performance of single-scale models using either long-scale (100\,ms) or short-scale (2\,ms) inputs. 
Tables~\ref{table:long_scale_performance} and~\ref{table:short_scale_performance} summarize the quantitative results for different detector combinations and spectral channel configurations.
The performance of single-scale models shows a strong dependence on spectral resolution.
For both long-scale and short-scale data, the full 128-channel configurations achieve the highest accuracy ($\textgreater$ 93\%). 
The long-scale, dual-detector model (L\textsubscript{NaI-128C,BGO-128C}) attains 96.86\% accuracy, whereas the best-performing short-scale model (S\textsubscript{NaI-128C}) reaches 94.27\%. 
However, fusing data from both NaI and BGO detectors provides only a marginal benefit for long-scale analysis and no clear advantage for short-scale signals, despite approximately doubling the computational cost. 
These results establish that high spectral resolution is critical for accurate classification, whereas the utility of multi-detector data is scale-dependent.

\subsection{Dual-Scale Fusion}
\label{subsec:results_fusion}
We combine long- and short-scale features using the proposed Cross-Scale Feature Fusion (CSFF) module to exploit their complementary temporal information. 
Table~\ref{table:fusion_methods} shows that CSFF consistently outperforms simple concatenation and element-wise addition, with the joint scale- and element-level attention achieving the best overall performance.
Across all spectral channel configurations, the resulting dual-scale models (Table~\ref{table:two_scale_fused_performance}) yield further accuracy gains over their single-scale counterparts. 
The optimal model \texttt{F\textsubscript{NaI-128C,BGO-128C}} achieves a peak accuracy of 97.23\%. 
Notably, strong performance is retained even when only three broad energy channels per detector are used:
the model \texttt{F\textsubscript{NaI-3C,BGO-3C}} achieves an accuracy of 95.50\% while maintaining a compact parameter size and low computational cost. 
This result indicates that high-fidelity classification can be achieved using standard, physically motivated few-channel light curves.

\subsection{Class-wise Behavior and Deployment Test}
\label{subsec:results_cm}
Figure~\ref{fig:confusion_matrices} compares the confusion matrices for the best-performing long-scale, short-scale, and dual-scale models.
The overall recall rate exceeded 90\%, and the rates for GRB, SFLARE and TGF even surpassing 98\%. 
The confusion matrix of the fused model exhibits the strongest diagonal dominance, as evidenced by the darker blue hues concentrated along the main diagonal. 
This indicates a significant reduction in inter-class confusion—particularly between traditionally ambiguous categories such as GRBs and particle-induced events—compared to the single-scale models.
The superior performance of the dual-scale architecture underscores the benefit of integrating both long-scale temporal profiles and short-scale variability for high-fidelity trigger classification.

Figure~\ref{fig:test_classify_trigs} compares the performance of the standard \textit{Fermi}/GBM in-flight trigger classification algorithm against our optimal model on three months (from 01 September 2025 to 30 November 2025) of real \textit{Fermi}/GBM trigger data. 
A key difference is that, due to the lack of reliable localization information, the GBM in-flight algorithm fails to correctly identify a significant number of solar flares (27 out of 80) and local particle events (21 out of 33). 
In contrast, our model achieves high-precision classification across all event categories.  
Our model shows high-precision recognition for each trigger category of events. 
These results demonstrate the robustness and practical utility of the proposed framework, which performs well on both curated datasets and real, unseen observational data.

\subsection{UMAP Visualization of the Learned Feature Space}
\label{subsec:umap_viz}
Using UMAP, Figure~\ref{fig:down_dim} visualizes the intrinsic structure of the feature representations learned by the optimal dual-scale model \(\text{F}_{\text{NaI-128C},\text{BGO-128C}}\).
The two-dimensional projection reveals a well-organized clustering topology, where triggers are grouped according to their predicted physical class: GRB, SFLARE, SGR, TGF, and PARTICLE events form distinct, separable clusters. 
Notably, events that the \textit{Fermi}/GBM pipeline failed to classify confidently (labeled \texttt{UNCERT}) predominantly occupy the boundary regions between these major clusters rather than forming an isolated group.
This spatial distribution indicates that the model’s latent space organizes events along a continuous morphological manifold, where ambiguous triggers naturally reside in intermediate positions that share characteristics with multiple canonical classes. 
The clear separation of primary classes, coupled with the boundary-localized uncertain events, visually corroborates the model’s ability to extract physically meaningful discriminative features and suggests that the \texttt{UNCERT} category largely consists of morphologically transitional or hybrid events rather than entirely novel phenomena.

\section{Discussion}
\label{sec:discussion}
Rapid and accurate classification of high-energy transient triggers is essential for modern time-domain and multi-messenger astrophysics. 
Our framework provides the low-latency identification required to initiate timely follow-up observations, enhances catalog purity by filtering out instrumental background, and supports population studies that constrain the physics of cosmic transients \citep{Fermi_GBM_catalog_fourth}.

\subsection{A General Framework for Trigger Classification}
We present a deep learning framework for the end-to-end classification of high-energy transient triggers.
The model operates directly on multi-band, background-preserved light curves. 
It eliminates the reliance on manually engineered features—such as hardness ratios and source localization—that are sensitive to observational conditions. 
The model is trained and validated on a curated dataset (Table~\ref{table:dataset}), which includes both astrophysical transients (e.g., GRBs, solar flares, SGRs, TGFs) and non-astrophysical, particle-induced backgrounds. 
This morphological diversity (Figure~\ref{fig:trig_samples}) highlights the core challenge of building a classifier that remains robust across variable signal-to-noise ratios, temporal profiles, and spectral evolutions.

\subsection{The Rationale for Dual-Scale Temporal Modeling}
Figure~\ref{fig:model} illustrates the dual-scale architecture central to our approach. The model processes long- and short-scale representations in parallel, capturing both extended morphology and rapid variability. Single-scale long models summarize broad temporal context but can smooth over short-duration features (Table \ref{table:long_scale_performance}). 
Single-scale short models resolve fine structure but lack broader context (Table \ref{table:short_scale_performance}). 
This dichotomy is especially relevant given the distinct physical origins of the transient classes: (1) cosmological GRBs span a wide range of durations and exhibit complex temporal structure \citep{GRB_duration_Kouveliotou_1993apjl, GRB_duration_ZhangBing_2004, GRB_type_IL_wang_2025, GRB_type_IL_tan_2025}; (2) magnetar-related SGRs are characterized by clustered short bursts \citep{Magnetars_Kaspi_2017AA, SGR_xie_2025, SGR_wang_2025}; (3) TGFs are ultra-short, high-energy atmospheric flashes \citep{TGF_Fishman_1994Science, TGF_GECAM_Zhao_2023, TGF_HXMT_Yi_2025SCPMA, TGF_Yi_2025SCPMA}; and (4) solar flares typically display smoother, extended emission profiles \citep{SFLARE_Fletcher_2011ssr}. 

\subsection{Adaptive Multi-Scale Fusion via the CSFF Module}
To reconcile these complementary representations, we propose a Cross-Scale Feature Fusion (CSFF) module. Unlike fixed strategies like concatenation, CSFF uses learned attention to balance contributions from each scale conditioned on the input (Table~\ref{table:fusion_methods}). 
This adaptive mechanism is critical, as it allows the network to dynamically emphasize long-scale patterns for temporally extended events and short-scale features for impulsive transients.
This adaptive fusion yields statistically significant performance gains over simple baseline fusion methods (e.g., concatenation). 

\subsection{State-of-the-Art Performance and Practical Efficiency}
Our dual-scale architecture achieves a state-of-the-art accuracy of 97.23\% (model \texttt{F\textsubscript{NaI-128C,BGO-128C}}; Table~\ref{table:two_scale_fused_performance}), representing a significant advance over existing methods.
For instance, the DL approach of \citet{trig_classify_DL_1} achieves 93\% accuracy but requires seven time-scale background-subtracted light curves. 
In contrast, our end-to-end framework attains higher accuracy using only 
two (or even one) temporal scales of raw light curves. 
It eliminates dependence on instrument-specific background modeling—a process prone to errors under highly variable conditions (e.g., solar flares).

A pivotal finding is that high accuracy is maintained with highly efficient configurations. 
The model \texttt{F\textsubscript{NaI-3C}}, which processes only three broad energy bands (5–100, 100–300, 300–900 keV), attains 95.07\% accuracy with ~845k parameters and 377 MFLOPs. This demonstrates that physically motivated, coarse spectral binning—combined with adaptive temporal fusion—preserves sufficient information for robust classification.

The critical role of temporal morphology is underscored by the \texttt{F\textsubscript{NaI-1C}} model, which achieves nearly 90\% accuracy using only a single energy band. 
This contrasts with traditional pipelines, whose underperformance may be attributed to their inability to adequately capture these complex morphological patterns \citep{trig_classify_1,trig_classify_gecam,trig_classify_3}. 

\subsection{Superiority Over In-Flight Algorithms and Robustness}
\begin{figure*}[h]
\centering
\begin{minipage}{0.98\textwidth}
    \includegraphics[width=0.371\linewidth]{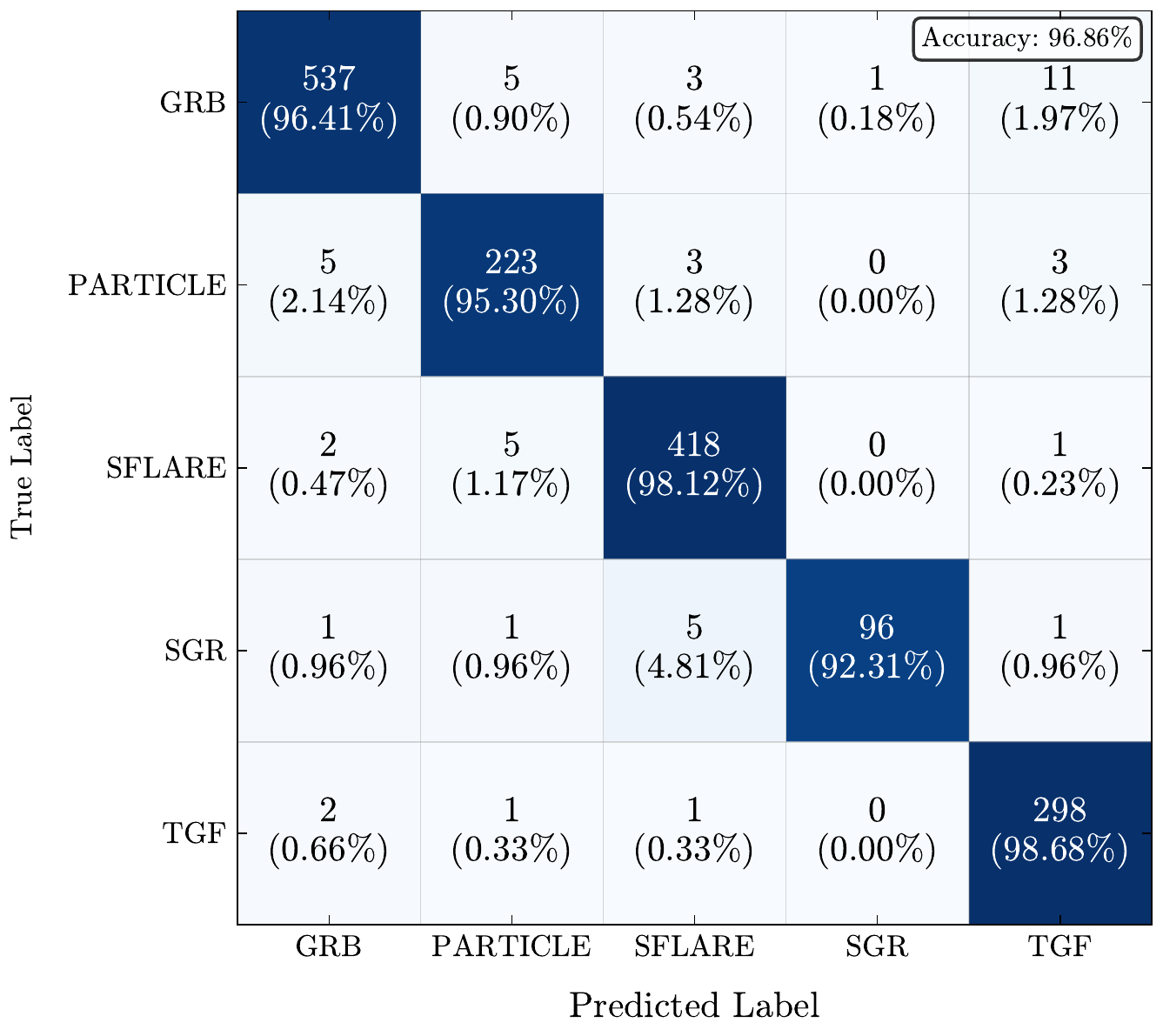}
    \includegraphics[width=0.30\linewidth]{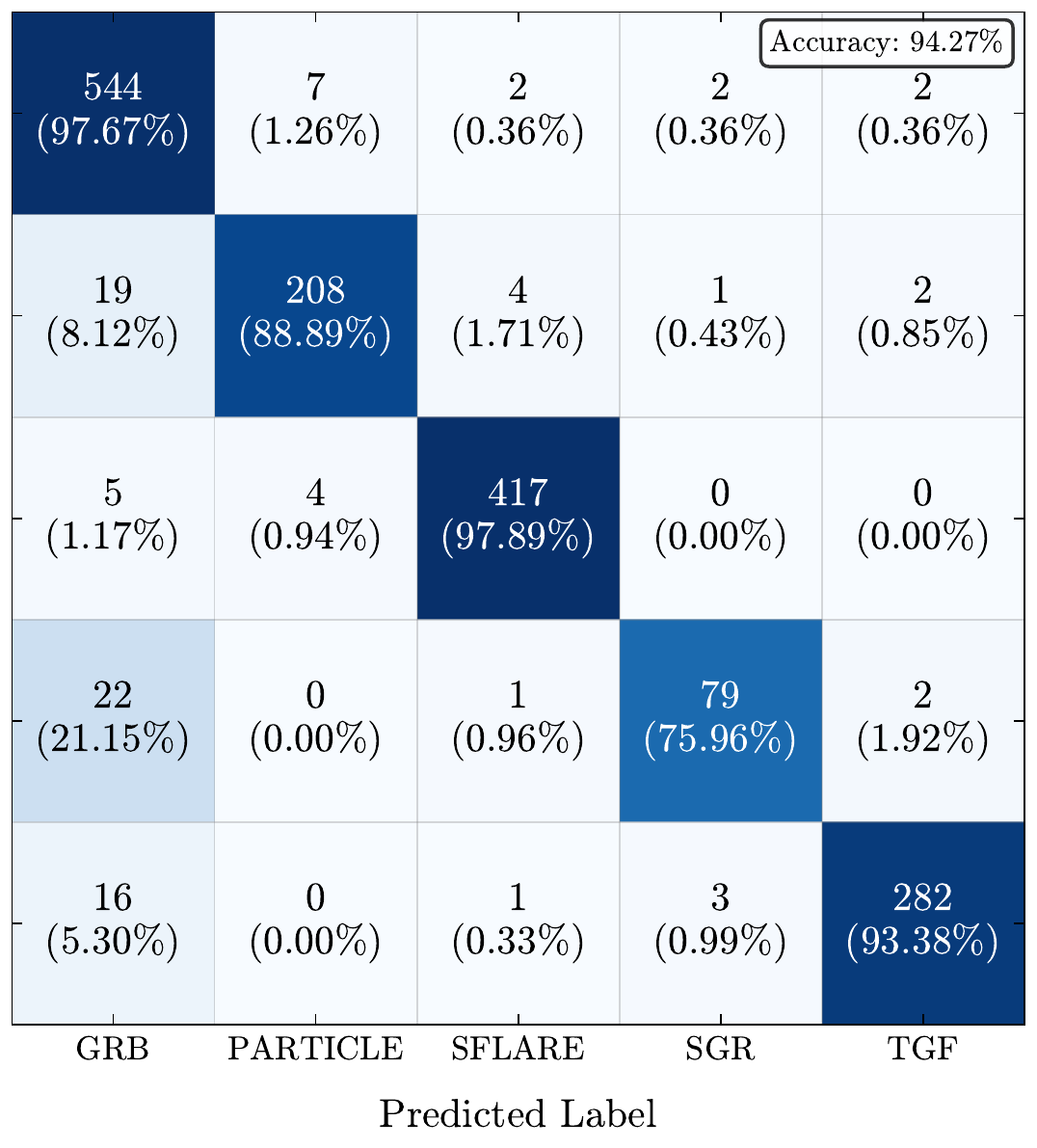}
    \includegraphics[width=0.30\linewidth]{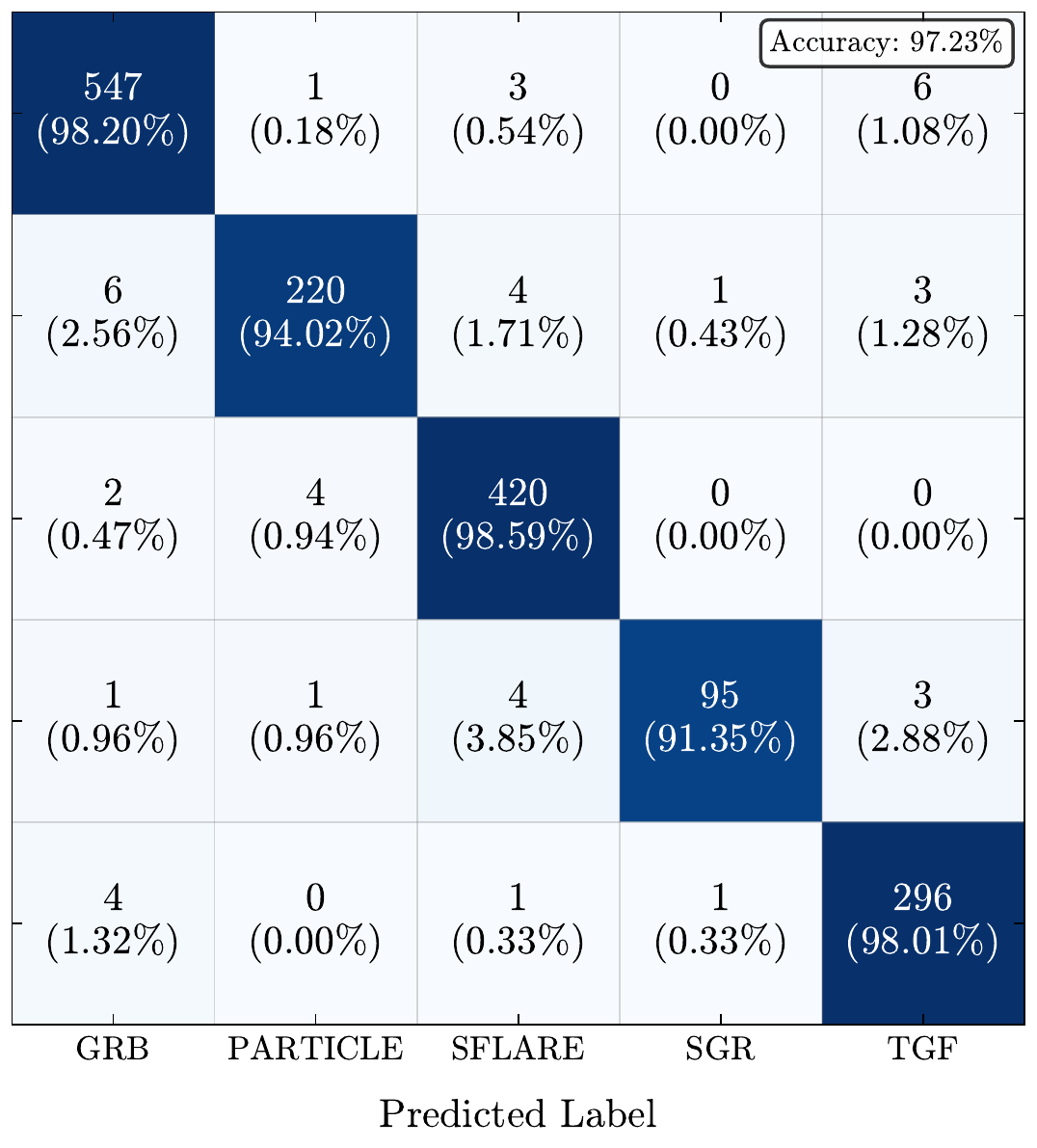}
\end{minipage}
\caption{
Confusion matrices for the best-performing models under three architectural settings:
the long-scale model \texttt{L\textsubscript{NaI-128C,BGO-128C}} (left),
the short-scale model \texttt{S\textsubscript{NaI-128C}} (middle),
and the dual-scale fused model \texttt{F\textsubscript{NaI-128C,BGO-128C}} (right).
Rows correspond to the true trigger classes and columns to the model-predicted classes.
Each cell reports the number of events, with the corresponding fraction relative to the true class shown in parentheses.
The overall classification accuracy for each model is indicated in the upper-right corner of each panel.
}
\label{fig:confusion_matrices}
\end{figure*}

\begin{figure*}[htbp]
\centering
\begin{minipage}{1\textwidth}
    \centering
    \includegraphics[width=0.44\linewidth]{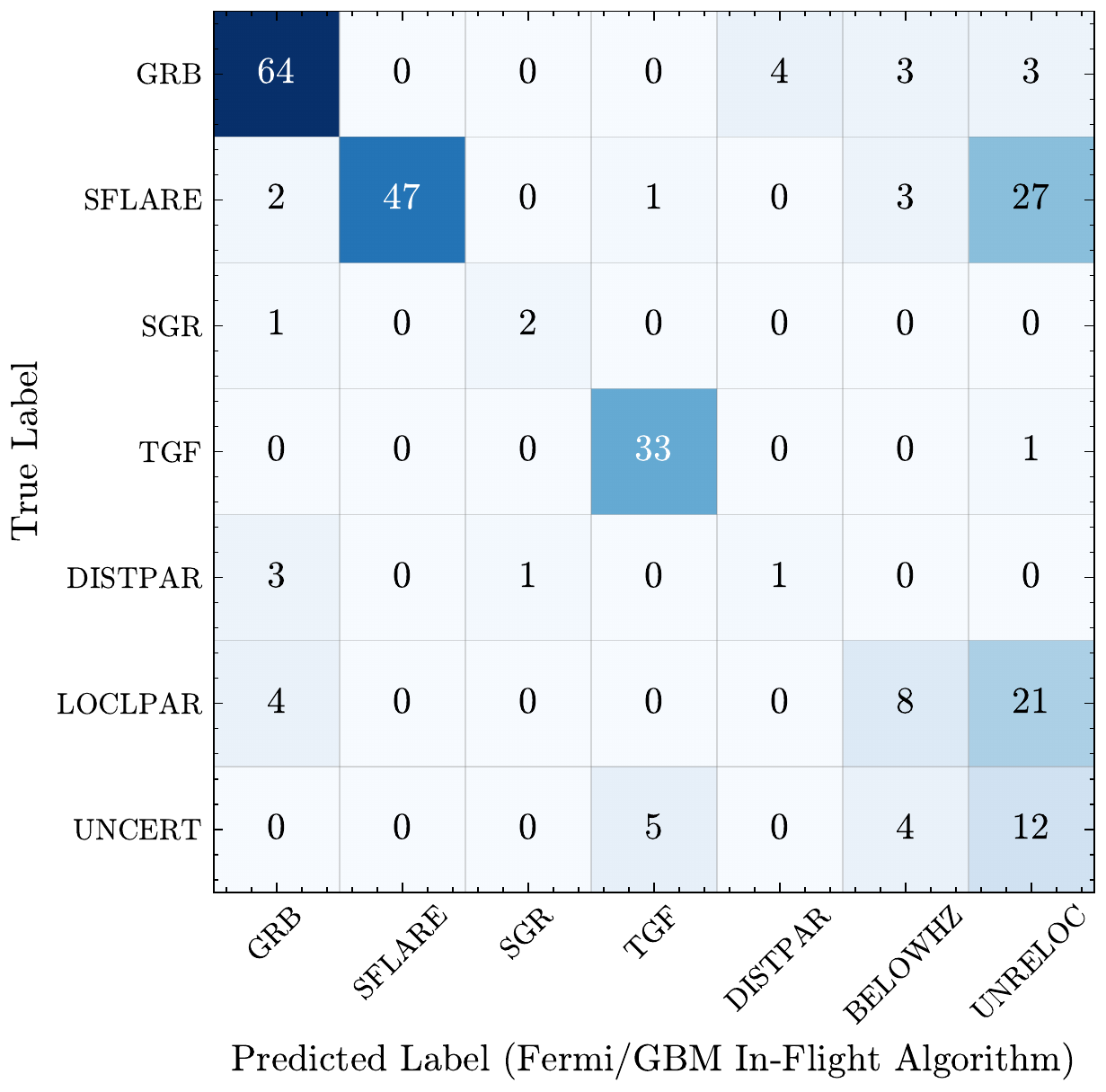}
    \hspace{3em}
    \includegraphics[width=0.338\linewidth]{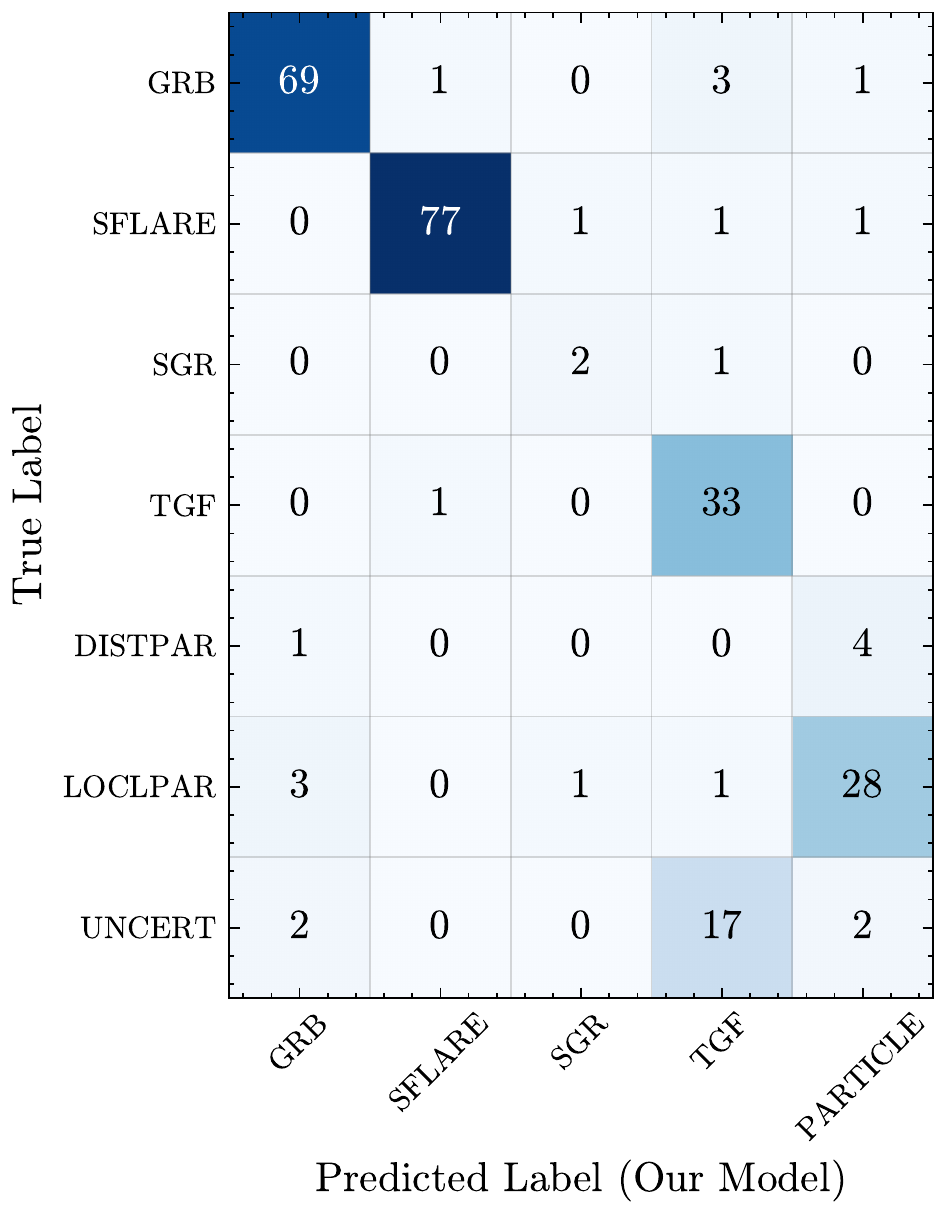}
\end{minipage}
\caption{
Confusion matrices for the three-month trigger dataset, contrasting predictions of the standard in-flight trigger classification algorithm (left) and our optimal dual-scale model (right). 
Rows represent the ground-truth labels from the GBM catalog, and columns represent the predicted classifications. 
The results of the in-flight trigger classification are derived from the records in the `Trigdat' file released by GBM. 
The \texttt{UNCERT} (truth labels) correspond to triggers that could not be confidently classified by the standard \textit{Fermi}/GBM pipeline. 
The classes \texttt{BELOWHZ} and \texttt{UNRELOC} denote ``Source below the horizon" and ``Unreliable location", respectively. 
Cell entries indicate the count of triggers for each true–predicted class pair. 
}
\label{fig:test_classify_trigs}
\end{figure*}

Class-wise performance, as visualized in the confusion matrices (Figure~\ref{fig:confusion_matrices}), reveals that the fused model reduces inter-class ambiguity compared to single-scale models, particularly for traditionally challenging distinctions such as short GRBs versus SGR events. 
The pronounced diagonal dominance underscores the model’s capability for high-confidence predictions. When evaluated on three months of archival \textit{Fermi}/GBM triggers, our framework achieves significantly higher accuracy than the operational in-flight trigger classifier (Figure~\ref{fig:test_classify_trigs}). 
This improvement is attained without reliance on reliable real-time localization—a notable advantage, given that precise onboard localization is often computationally prohibitive. 
Moreover, the model remains robust when background estimation is compromised, such as during solar flares. 
For practical deployment on satellites or in ground‑based pipelines, an ``uncertain'' classification category remains essential. 
This category can accommodate ambiguous cases—including instrumental anomalies, currently unknown astrophysical phenomena, or events caused by known sources under occultation (e.g., earth occultation). 
By applying confidence thresholds to the model’s predictions, low‑certainty triggers can be automatically assigned to this category, enabling human review or further analysis while preserving the integrity of automated processing.

This combination of high accuracy, low computational cost, and instrument-agnostic inputs establishes a practical pathway for real-time applications. 
For instance, upon detecting a trigger, satellites like SVOM could transmit compact three-band light curves to the ground in near real-time \citep{SVOM_GRM} for immediate classification, which would facilitate timely follow-up observations. 
Furthermore, the minimal computational footprint makes the framework particularly suited for in-flight processing under stringent resource constraints. This aligns with the growing trend of embedding machine learning directly on spacecraft to enable autonomous data screening and rapid event identification \citep{AI_Onboard_Satellite_1, AI_Onboard_Satellite_2, AI_Onboard_Satellite_3}.

\subsection{Interpretability of the Learned Feature Space}
\begin{figure*}[htbp]
\centering
\includegraphics[width=0.75\textwidth]{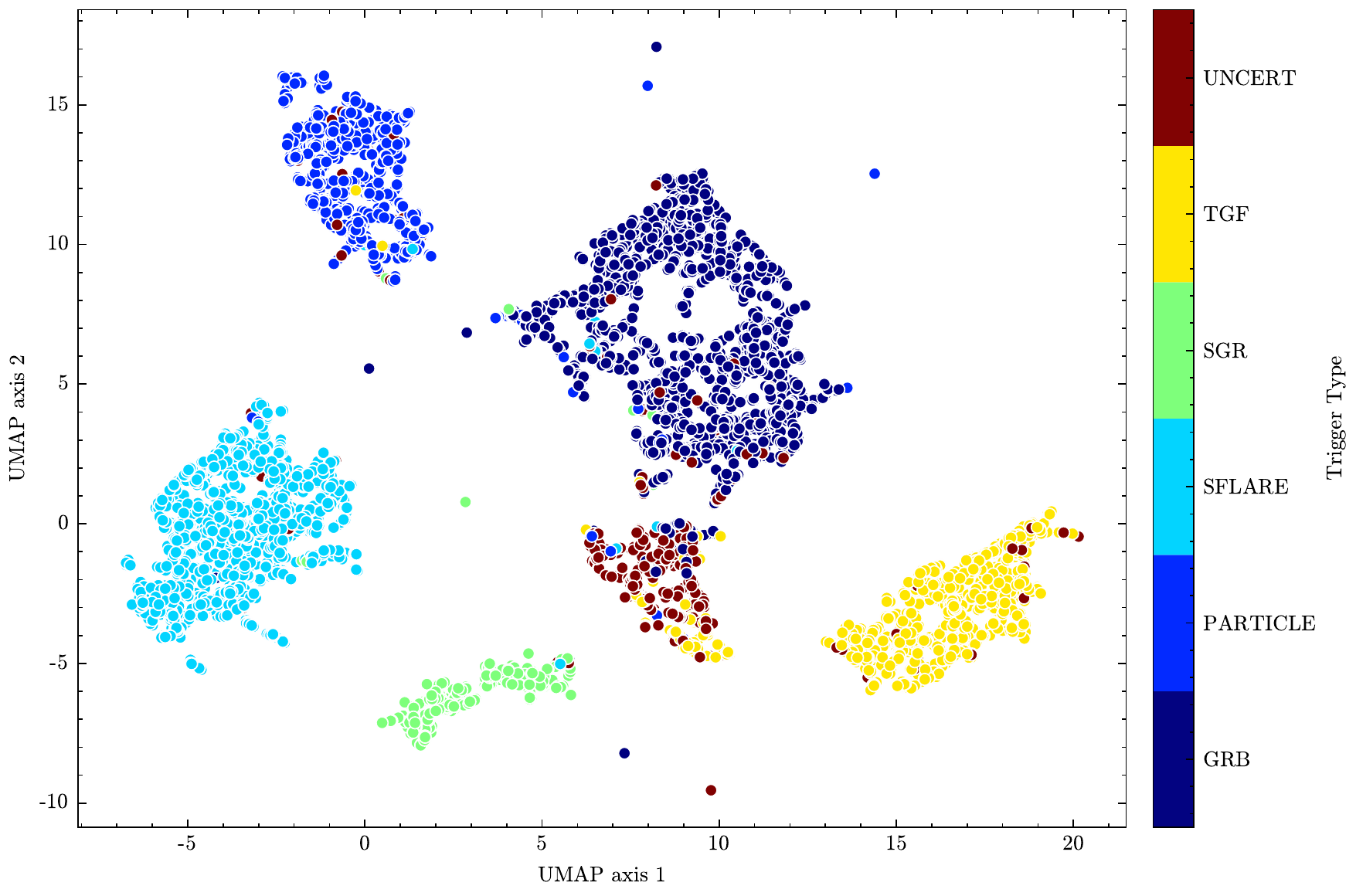}
\caption{
UMAP visualization of the feature space learned by the optimal dual-scale model \texttt{F\textsubscript{NaI-128C,BGO-128C}}.
Features are extracted from all events in the dataset and projected onto two dimensions using UMAP.
Each point represents a single trigger and is colored by its model-predicted class: GRB, SFLARE, SGR, TGF, and PARTICLE.
Events labeled as \texttt{UNCERT} (red) correspond to triggers that could not be confidently classified by the standard \textit{Fermi}/GBM pipeline.
The projection illustrates the clustering structure of the learned representations and the relative distribution of uncertain events with respect to confidently classified trigger populations.
}
\label{fig:down_dim}
\end{figure*}

We projected the high-dimensional features from the optimal dual-scale model onto a two-dimensional manifold using UMAP (Figure~\ref{fig:down_dim}). 
The projection reveals well-separated clusters corresponding to major physical classes (e.g., GRBs, SGRs).
Uncertain events (\texttt{UNCERT}) predominantly occupy boundary regions between clusters. This indicates the model’s latent space not only separates well-characterized events but also organizes them along a continuous morphological manifold. 
Ambiguous triggers reside in intermediate positions, suggesting they exhibit hybrid characteristics shared by multiple canonical classes. 
This implies the \texttt{UNCERT} category may comprise morphologically transitional events or fundamentally novel phenomena. 
Manual inspection of events in the \texttt{UNCERT} category revealed instances with hybrid features, some of which are suspected candidates for known classes like GRBs or TGFs that require multi-satellite confirmation. 
These candidates usually require joint analysis of multiple satellites for further confirmation. 
A deeper analysis of these ambiguous candidates, potentially through multi-satellite joint analysis, is planned for future work.

Complementing this, Grad-CAM probes confirm the model consistently attends to physically salient regions in the time–energy domain. 
For representative triggers (Figure~\ref{fig:trig_samples}), heatmaps highlight specific temporal and spectral bins that align with known phenomenological signatures—such as brief, high-energy peaks for TGFs or smooth emission for solar flares. 
This correspondence clarifies the decision mechanism and strengthens confidence in the model’s physical consistency.

\subsection{Limitations and Future Directions}
The performance of the model is inherently bounded by the representativeness of the training data. Residual misclassifications between morphologically similar event classes indicate that integrating auxiliary observational information—such as coarse source localization, multi-instrument coincidence signals, or contextual metadata—could further improve discriminative power. 
Additionally, the selection of optimal temporal scales and the handling of low signal‑to‑noise ratio events warrant more systematic investigation in future work.

As observational data volumes grow, semi‑supervised or unsupervised learning techniques may offer a promising pathway for identifying rare or previously unmodeled transient phenomena. The instrument‑agnostic design of the framework ensures its applicability to both current and forthcoming high‑energy astronomy missions, positioning it as a potential core component in next‑generation automated transient analysis pipelines \citep{trig_search_ETJASMIN_I, BREAKFAST_pipline_wang_2025arxiv}.

\section{Summary}
\label{sec:summary}

In this work, we have addressed the critical need for rapid and accurate high-energy transient classification in modern time-domain astronomy. 
We present \texttt{LUNCH}, a deep learning framework that directly processes raw, multi-band light curves.
This approach bypasses the dependency on handcrafted features (such as hardness ratios, which require background subtraction) and on reliable real-time localization—both of which are limiting factors for conventional classification methods. 

Our framework employs a dual-scale temporal modeling architecture and has been validated on a comprehensive dataset spanning 15 years of \textit{Fermi}/GBM observations. 
The key findings are as follows: 

1.  \textit{State-of-the-Art Performance:} The optimal dual-scale fused model (\texttt{F\textsubscript{NaI-128C,BGO-128C}}) achieves a state-of-the-art classification accuracy of 97.23\%, surpassing both single-scale models. 
This is reflected in the strong diagonal dominance of its confusion matrix, indicating a significant reduction in inter-class ambiguity.

2.  \textit{Efficiency and Practicality:} A highly efficient variant of the model, operating on only three broad energy bands (\texttt{F\textsubscript{NaI-3C}}), retains 95.07\% accuracy with a minimal computational footprint. This demonstrates that physically motivated, coarse spectral information, when combined with adaptive temporal fusion, provides a robust basis for classification, enabling real-time or in-flight deployment. 

3.  \textit{Robustness and Interpretability:} The framework is robust under challenging conditions where traditional methods struggle, such as during periods of intense and variable background activity (e.g., solar flares). 
Interpretability analyses confirm its physical consistency. UMAP visualization shows that the learned feature space forms well-separated clusters corresponding to distinct event classes, with ambiguous events naturally populating the boundaries—indicating a morphologically continuous latent manifold.
Furthermore, Grad-CAM visualizations show that the model consistently focuses on physically salient time-energy regions during predictions.

4.  \textit{Validation Against Operational Systems:} Evaluated on three months of independent, archival \textit{Fermi}/GBM trigger data, our framework achieves significantly higher classification accuracy than the operational in-flight trigger classifier, all without requiring reliable localization.

Our work establishes that the discriminative signatures for high-energy transients are intrinsically encoded in the temporal morphology and spectral evolution across energy bands. By integrating this insight into an efficient, end-to-end, and interpretable deep learning architecture, we provide a practical solution for automated trigger classification. This framework is instrument-agnostic and is poised to enhance the scientific return of current and future high-energy missions by enabling low-latency, high-fidelity event screening, directly supporting rapid multi-wavelength and multi-messenger follow-up in the era of time-domain astronomy.

\section*{Acknowledgements}
This work is supported by 
the National Key R\&D Program of China (Grant No. 2021YFA0718500), 
the Strategic Priority Research Program of the Chinese Academy of Sciences (Grant No. 
XDB0550300), 
the National Natural Science Foundation of China 
(Grant No. 12273042, 
12494572
),
the National Natural Science Foundation of China (41827807 and 61271351), 
the Science and Technology Innovation Plan of Shanghai Science and Technology Commission
(22DZ1209500). 
We appreciate the public data of Fermi/GBM.
The GECAM (Huairou-1) mission is supported by the Strategic Priority Research Program on Space Science of the Chinese Academy of Sciences (XDA15360000).

\bibliographystyle{aasjournal}
\bibliography{paper}

\begin{deluxetable*}{lccccccccc}[h]
\label{table:long_scale_performance}
\tablecaption{Performance and Complexity Comparison on Long-Scale (100 ms) Data}
\tablehead{
\colhead{Model ID} & \colhead{Detector} & \colhead{Channels} & \colhead{Architecture} & \colhead{Accuracy} & \colhead{Precision} & \colhead{Recall} & \colhead{F1} & \colhead{Params} & \colhead{FLOPs} \\
& & & & \colhead{(\%)} & \colhead{(\%)} & \colhead{(\%)} & \colhead{(\%)} & \colhead{(k)} & \colhead{(M)}
}
\startdata
\multicolumn{10}{l}{\textbf{NaI Detector Configurations}} \\
L\textsubscript{NaI-1C}   & NaI & 1    & Single-ResNet\textsubscript{32,8}   & 82.99 & 80.46 & 82.28 & 81.13 & 106.23 & 22.30 \\
L\textsubscript{NaI-3C}   & NaI & 3    & Single-ResNet\textsubscript{64,64}  & 93.47 & 93.29 & 92.77 & 92.97 & 425.22 & 125.71 \\
L\textsubscript{NaI-128C} & NaI & 128  & Single-ResNet\textsubscript{64,64}  & 96.67 & 96.72 & \textbf{96.25} & \textbf{96.46} & 425.22 & 3617.14 \\
\cmidrule(l){1-10}
\multicolumn{10}{l}{\textbf{NaI+BGO Fused Configurations}} \\
L\textsubscript{NaI-1C,BGO-1C}     & NaI+BGO & 1+1   & Single-ResNet\textsubscript{32,8}   & 87.12 & 85.60 & 87.20 & 86.30 & 106.23 & 22.30 \\
L\textsubscript{NaI-3C,BGO-3C}     & NaI+BGO & 3+3   & Single-ResNet\textsubscript{32,16}  & 93.22 & 92.66 & 92.51 & 92.57 & 106.53 & 48.70 \\
L\textsubscript{NaI-128C,BGO-128C} & NaI+BGO & 128+128 & Single-ResNet\textsubscript{64,64}  & \textbf{96.86} & \textbf{96.83} & 96.16 & \textbf{96.46} & 425.22 & 7234.27 \\
\enddata
\tablecomments{
Performance and complexity metrics for single-scale (long) models. \\
\textbf{Notation:} L = Long-scale (100 ms resolution). Channel subscripts: 1 (summed), 3 (energy-segmented: NaI: 25-100/100-300/300-900 keV; BGO: 300-900/900-2000/2000-40000 keV), 128 (complete spectral channels). Comma (,) separates multi-detector configurations. \\
\textbf{Architecture:} Subscript in model name (e.g., \textsubscript{32,8}) denotes convolutional filters and FC layer neurons. \\
\textbf{Metrics:} All performance values are percentages (\%). Params in thousands (k), FLOPs in millions (M) for a single forward pass.
}
\end{deluxetable*}

\begin{deluxetable*}{lccccccccc}[h]
\label{table:short_scale_performance}
\tablecaption{Performance and Complexity Comparison on Short-Scale (2 ms) Data}
\tablehead{
\colhead{Model ID} & \colhead{Detector} & \colhead{Channels} & \colhead{Architecture} & \colhead{Accuracy} & \colhead{Precision} & \colhead{Recall} & \colhead{F1} & \colhead{Params} & \colhead{FLOPs} \\
& & & & \colhead{(\%)} & \colhead{(\%)} & \colhead{(\%)} & \colhead{(\%)} & \colhead{(k)} & \colhead{(M)}
}
\startdata
\multicolumn{10}{l}{\textbf{NaI Detector Configurations}} \\
S\textsubscript{NaI-1C}   & NaI & 1    & Single-ResNet\textsubscript{16,8}  & 68.82 & 69.25 & 64.97 & 62.19 & 27.03 & 11.40 \\
S\textsubscript{NaI-3C}   & NaI & 3    & Single-ResNet\textsubscript{16,16} & 75.79 & 79.90 & 74.59 & 75.81 & 27.21 & 16.36 \\
S\textsubscript{NaI-128C} & NaI & 128  & Single-ResNet\textsubscript{64,32} & \textbf{94.27} & \textbf{94.74} & 90.76 & \textbf{92.52} & 422.98 & 7233.35 \\
\cmidrule(l){1-10}
\multicolumn{10}{l}{\textbf{NaI+BGO Fused Configurations}} \\
S\textsubscript{NaI-1C,BGO-1C}     & NaI+BGO & 1+1   & Single-ResNet\textsubscript{16,32} & 72.09 & 63.14 & 66.46 & 64.06 & 27.56 & 11.40 \\
S\textsubscript{NaI-3C,BGO-3C}     & NaI+BGO & 3+3   & Single-ResNet\textsubscript{16,64} & 80.41 & 83.50 & 80.91 & 82.07 & 28.26 & 25.00 \\
S\textsubscript{NaI-128C,BGO-128C} & NaI+BGO & 128+128 & Single-ResNet\textsubscript{64,64} & 93.96 & 94.06 & \textbf{90.99} & 92.38 & 425.22 & 14466.69 \\
\enddata
\tablecomments{
Performance and complexity metrics for single-scale (short) models. \\
\textbf{Notation:} S = Short-scale (2 ms resolution). Channel subscripts: 1 (summed), 3 (energy-segmented: NaI: 25-100/100-300/300-900 keV; BGO: 300-900/900-2000/2000-40000 keV), 128 (complete spectral channels). Comma (,) separates multi-detector configurations. \\
\textbf{Architecture:} Subscript in model name (e.g., \textsubscript{16,8}) denotes convolutional filters and FC layer neurons. \\
\textbf{Metrics:} All performance values are percentages (\%). Params in thousands (k), FLOPs in millions (M) for a single forward pass. 
}
\end{deluxetable*}

\begin{deluxetable*}{lcccccc}[h]
\label{table:fusion_methods}
\tablecaption{Performance and Complexity Comparison of Multi-Scale Feature Fusion Methods}
\tablehead{
\colhead{Fusion Method} & \colhead{Accuracy} & \colhead{Precision} & \colhead{Recall} & \colhead{F1-Score} & \colhead{Params} & \colhead{FLOPs} \\
& \colhead{(\%)} & \colhead{(\%)} & \colhead{(\%)} & \colhead{(\%)} & \colhead{(k)} & \colhead{(M)}
}
\startdata
Concatenation & 94.14 & 94.28 & 92.19 & 93.15 & 842.55 & 576.31 \\
Element-wise Addition & 94.64 & 94.74 & 93.68 & 94.18 & 842.04 & 576.31 \\
\midrule
CSFF ($\alpha$ only) & 95.26 & 95.26 & 94.48 & 94.86 & 842.06 & 576.31 \\
CSFF ($\beta$ only) & 95.13 & 95.09 & 93.94 & 94.47 & 842.06 & 576.31 \\
\textbf{CSFF ($\alpha$ \& $\beta$)} & \textbf{95.50} & \textbf{95.32} & \textbf{94.74} & \textbf{95.03} & 842.06 & 576.31 \\
\enddata
\tablecomments{
Comparison of different feature fusion strategies for combining long-scale and short-scale encoder outputs (The input data of the current experimental model is \texttt{NaI-3C,BGO-3C}.). CSFF denotes our proposed Cross-Scale Feature Fusion Module. The terms $\alpha$ and $\beta$ represent the scale-level and element-level attention mechanisms, respectively, as defined in Section~\ref{subsec:fusion_module}. Params (in thousands) counts the number of trainable parameters; FLOPs (in millions) estimates the floating-point operations required for a single forward pass. 
}
\end{deluxetable*}

\begin{deluxetable*}{lccccccccc}[h]
\label{table:two_scale_fused_performance}
\tablecaption{Performance and Complexity of Two-Scale Fused Models}
\tablehead{
\colhead{Model ID} & \colhead{Detector} & \colhead{Channels} & \colhead{Architecture} & \colhead{Accuracy} & \colhead{Precision} & \colhead{Recall} & \colhead{F1} & \colhead{Params} & \colhead{FLOPs} \\
& & & & \colhead{(\%)} & \colhead{(\%)} & \colhead{(\%)} & \colhead{(\%)} & \colhead{(k)} & \colhead{(M)}
}
\startdata
\multicolumn{10}{l}{\textbf{NaI Detector Configurations}} \\
F\textsubscript{NaI-1C}   & NaI & 1\textbar1    & Dual-ResNet\textsubscript{64,64} & 88.91 & 87.26 & 86.34 & 86.74 & 845.98 & 264.32 \\
F\textsubscript{NaI-3C}   & NaI & 3\textbar3    & Dual-ResNet\textsubscript{64,64} & 95.07 & 95.28 & 93.85 & 94.51 & 845.98 & 377.02 \\
F\textsubscript{NaI-128C} & NaI & 128\textbar128 & Dual-ResNet\textsubscript{64,8}  & 97.17 & \textbf{97.73} & \textbf{96.32} & \textbf{97.00} & 842.06 & 10850.48 \\
\cmidrule(l){1-10}
\multicolumn{10}{l}{\textbf{NaI+BGO Fused Configurations}} \\
F\textsubscript{NaI-1C,BGO-1C}     & NaI+BGO & (1+1)\textbar(1+1)     & Dual-ResNet\textsubscript{64,64}  & 92.36 & 91.60 & 90.73 & 91.12 & 845.98 & 264.32 \\
F\textsubscript{NaI-3C,BGO-3C}     & NaI+BGO & (3+3)\textbar(3+3)     & Dual-ResNet\textsubscript{64,16}  & 95.50 & 95.33 & 94.75 & 95.03 & 842.62 & 576.31 \\
F\textsubscript{NaI-128C,BGO-128C} & NaI+BGO & (128+128)\textbar(128+128) & Dual-ResNet\textsubscript{64,8}  & \textbf{97.23} & 97.26 & 96.04 & 96.61 & 842.06 & 21700.96 \\
\enddata
\tablecomments{
Performance and complexity metrics for dual-scale fused models.\\
\textbf{Notation:} F = Fused two-scale. Channel format: long-scale\,\textbar\,short-scale; plus sign (+) denotes detector channel concatenation; comma (,) separates multi-detector configurations. Channel subscripts: 1 (summed), 3 (energy-segmented), 128 (complete spectral channels). \\
\textbf{Architecture:} Subscript in model name (e.g., \textsubscript{64,8}) denotes convolutional filters and FC layer neurons. \\
\textbf{Metrics:} All performance values are percentages (\%). Params in thousands (k), FLOPs in millions (M) for a single forward pass. 
}
\end{deluxetable*}

\end{CJK*}
\end{document}